\documentclass[aps, prx, superscriptaddress, twocolumn, amssymb, longbibliography]{revtex4-2}

\usepackage{mathtools} 
\usepackage{graphicx} 
\usepackage{placeins} 
\usepackage{booktabs} 

\begin{document}


\title{Floquet Engineering Clock Transitions in Magnetic Molecules}

\author{Andrew Cupo}
\email{a.cupo@northeastern.edu}
\affiliation{Department of Physics, Northeastern University, Boston, Massachusetts, 02115, USA}

\author{Shuanglong Liu}
\affiliation{Department of Physics, Northeastern University, Boston, Massachusetts, 02115, USA}

\author{Silas Hoffman}
\affiliation{Laboratory for Physical Sciences, 8050 Greenmead Drive, College Park, Maryland 20740, USA}

\author{X.-G. Zhang}
\affiliation{Department of Physics and the Quantum Theory Project, University of Florida, Gainesville, Florida 32611, USA}

\author{Hai-Ping Cheng}
\email{ha.cheng@northeastern.edu}
\affiliation{Department of Physics, Northeastern University, Boston, Massachusetts, 02115, USA}


\begin{abstract}
We theoretically study Floquet engineering of magnetic molecules via a time-periodic magnetic field that couples to the emergent total electronic spin of the metal center. By focusing on the low-lying energy levels using an $S = 1$ spin Hamiltonian containing the zero-field and Zeeman terms, we demonstrate their continuous tunability under the Floquet field. Remarkably, under the action of linearly polarized Floquet controls, the energy levels of a clock transition qubit retain their stability against variations in an external static magnetic field. This property is closely linked to having a net-zero total Zeeman shift, which results from both static and effective dynamical contributions. Further, using second-order Van Vleck degenerate perturbation theory, we derived analytically an effective Hamiltonian, which explicitly shows the dependence of the renormalized zero-field tensor on the driving field. Based on our theoretical predictions, experimentalists will be able to dynamically tune qubit energy gaps to values that are useful in their specific laboratory settings, while retaining the spin decoherence suppressing effect of maintaining a clock transition.
\end{abstract}


\maketitle




\section{Introduction}
\label{sec:intro}

How to fully exploit the wide tunability of magnetic molecules to unlock their potential for quantum devices has been a challenge in materials physics \cite{graham2017forging, escalera2018spin, gaita2019molecular, atzori2019second, coronado2020molecular, aravena2020spin, wasielewski2020exploiting, carretta2021perspective, moreno2021measuring, chilton2022molecular, lunghi2022computational, forment2022chiral, moreno2023single, bernot2023get, fursina2023toward, chiesa2024molecular, gabarro2024challenges, chiesa2023quantum, moreno2024magnetic, guo2024understanding}. Typically, a magnetic molecule consists of a transition or lanthanide metal center with a surrounding ligand field containing mostly carbon, nitrogen, oxygen, and hydrogen. The metal center hosts an emergent total electronic spin $S$, leading to $2S+1$ low-lying energy levels, an isolated pair of which could function as a qubit \footnote{In some cases $S$ is not a good quantum number, but the total electronic angular momentum $J$ is}. Even at room temperature, a handful of these molecules exhibit useful spin coherence times on the order of 1 $\mu$s \cite{bader2014room, atzori2016quantum, amdur2022chemical, maylander2023room, mena2024room}. Additional enhancement of qubit spin coherence may be achieved by operation at a clock transition point, whereby the stability of each energy level against variations in an external static magnetic field leads to decoupling from the nuclear spin bath \cite{shiddiq2016enhancing, kundu2023electron, chen2025enhancing}.

Discussions on tuning molecular properties have predominantly centered on chemical modifications, which allow for discrete and time-fixed changes. In contrast, external-field-based approaches, such as Floquet engineering, have emerged as a promising avenue for driving molecular systems toward novel quantum phenomena. Notably, first-principles studies have proposed using Floquet driving fields to achieve Thouless pumping in molecular systems, highlighting the potential of dynamic control in this context \cite{zhou2021first, zhou2023molecular}. More generally, time-periodic modulation of a system Hamiltonian renormalizes observables of interest \cite{grifoni1998driven, basov2017towards, oka2019floquet, giovannini2020floquet, harper2020topology, rudner2020band, rodriguez2021low}. The natural question that arises is: Can Floquet engineering allow one to dynamically alter the properties of magnetic molecules, with continuous tunability enabled by varying control parameter(s)? 

Continuous tuning is essential for practical applications. For instance, the frequency range of existing electron paramagnetic resonance (EPR) equipment imposes constraints on the usable molecular clock transition gap. While chemical substitution can bring the gap energy closer to this range, Floquet engineering enables precise adjustments, allowing a broader selection of molecules to be studied using existing experimental setups. Furthermore, the ability to amplitude modulate the control on and off enables spin state initialization and the execution of quantum gates on nanosecond timescales \cite{castro2022optimal}.

In this paper, we use a model electronic spin Hamiltonian to study the possibility of tuning the low-lying energy levels of magnetic molecules via a Floquet drive, a time-periodic magnetic field in the few GHz regime with continuously adjustable amplitude that couples to the emergent total electronic spin $S$ originating from the molecular metal center. To elaborate further, our approach encompasses three key characteristics: \textit{continuous}, \textit{ultrafast}, and \textit{reversible}. The differences between the renormalized energy levels will be \textit{continuously} tunable by varying the amplitude of the Floquet drive. In practice, the magnetic molecule begins at equilibrium and the Floquet control is amplitude modulated from zero up to a targeted value, after which it remains time-periodic for a specified duration. So long as the ramping time is at least a few driving cycles long (in our case about 1 ns), the predictions from Floquet theory are expected to be accurate, and the state preparation can be classified as \textit{ultrafast} \cite{d2015dynamical, castro2022floquet, lucchini2022controlling, cupo2024preparation}. Lastly, when the Floquet drive is amplitude modulated off, the molecule will revert to its equilibrium energy level structure, with some assistance from relaxation processes like spin-phonon coupling \cite{lunghi2023spin}. In this sense, the effect of the quantum control is \textit{reversible}, at least on the timescales associated with the relaxation processes. 

The outline of the paper is as follows. In Sec.~\ref{sec:theory} we lay out the mathematical theory that will be required to model Floquet-driven magnetic molecules. We first introduce the static electronic spin Hamiltonian as a starting point (\ref{sec:theory}A). Once the emergent total electronic spin $S$ is coupled to a time-periodic magnetic field, we provide a comprehensive overview of the Floquet formalism as it applies to the time-dependent Schr\"{o}dinger equation (\ref{sec:theory}B). The stability of the Floquet energy levels against variations in an external static magnetic field is assessed via a generalization of the Hellmann-Feynman theorem (\ref{sec:theory}C). Afterwards, an approach for computing the exact time-independent effective Hamiltonian that generates the one-cycle stroboscopic dynamics is described (\ref{sec:theory}D), and we define the related dynamical cancellation problem (\ref{sec:theory}E). Lastly, we derive an approximate effective Hamiltonian analytically via the second-order Van Vleck expansion (\ref{sec:theory}F). Moving on to Sec.~\ref{sec:results}, we describe the experimentally relevant molecular and control parameters utilized in the simulations. The core results and analysis of the paper consist of the energy level renormalization and corresponding stability determination (\ref{sec:results}A), a comparison to the dynamical cancellation problem (\ref{sec:results}B), and physical insight from the derived analytical effective Hamiltonian (\ref{sec:results}C). To support experimental realization, in Sec.~\ref{sec:results}D we present three approaches for implementing the Floquet drive, and provide practical parameter values for the most promising method based on existing work. Given that the ultimate application of our system is in quantum information science, to wrap up the discussion we discuss how state initialization, quantum gate application, and state readout relevant to quantum computation can be generalized from a standard qubit to a \textit{Floquet qubit} consisting of a pair of Floquet eigenstates corresponding to two non-degenerate quasienergy levels (\ref{sec:results}E) \cite{huang2021engineering, gandon2022engineering}. In Sec.~\ref{sec:conc} we conclude with a summary and provide an outlook with suggestions for future research directions.


\section{Theory of Floquet-Driven Magnetic Molecules}
\label{sec:theory}


\subsection{Static System}

We consider a magnetic molecule with a single metal center that is described accurately by an electronic spin Hamiltonian \cite{stoll2006easyspin} containing the zero-field ($S \geq 1$) and Zeeman terms
\begin{equation}
H_{\textrm{static}}
=
H_{\textrm{ZF}} + H_{\textrm{Zeeman}}
,
\label{ham_eq}
\end{equation}
\begin{equation}
H_{\textrm{ZF}}
=
\vec{s}\,^T \tilde{D} \vec{s}
=
D \left[ s_z^2 - \frac{1}{3} S (S+1) \boldsymbol 1 \right] + E ( s_x^2 - s_y^2 )
,
\label{ham_zf}
\end{equation}
\begin{equation}
H_{\textrm{Zeeman}}
=
\mu_B (\vec{B}_s)^T \tilde{g} \vec{s}
.
\label{ham_zeeman}
\end{equation}
With total spin $S$, let $\vec{S}$ be the vector of spin operators so that $\vec{s} = \vec{S}/\hbar$ is its dimensionless counterpart given as
\begin{equation}
\vec{s}
=
s_x \hat{x}
+
s_y \hat{y}
+
s_z \hat{z}
.
\label{spin_vector}
\end{equation}
In the zero-field term, the $3 \times 3$ $D$-tensor ($\tilde{D}$) is diagonal, and $x$, $y$, and $z$ are the principal axes of the molecule. Furthermore, for a traceless $\tilde{D}$, there are only two zero-field splitting parameters $D$ and $E$. Physically, the zero-field term describes the lifting of degeneracy in the ground state, as the total spin $S$ arises from the collective spins of the electrons within the many-body system. Note that since we will ultimately only apply our theory to $S = 1$, we do not require terms beyond quadratic order in the spin operators. Additionally, in the Zeeman term, the spin couples to a static external magnetic field $\vec{B}_s$ via the molecule-specific $3 \times 3$ $g$-tensor ($\tilde{g}$). The time-independent Schr\"{o}dinger equation (TISE) provides us with the energy levels for the total spin $S$ manifold
\begin{equation}
H_{\textrm{static}}
\varphi_n
=
E_n
\varphi_n
.
\label{tise}
\end{equation}


\subsection{Floquet Engineering}

Dynamical control is enabled by coupling the emergent total electronic spin to a time-dependent external magnetic field
\begin{equation}
H_{\textrm{TD}}(t)
=
\mu_B (\vec{B}(t))^T \tilde{g} \vec{s}
\label{ham_td}
\end{equation}
leading to the time-dependent Hamiltonian
\begin{equation}
H(t)
=
H_{\textrm{static}} + H_{\textrm{TD}}(t)
.
\label{ham_time}
\end{equation}
In particular we focus on time-periodic fields
\begin{equation}
\vec{B}(t+T) = \vec{B}(t)
\rightarrow
H(t+T) = H(t)
,
\label{time-periodic}
\end{equation}
which allows the time-dependent Schr\"{o}dinger equation (TDSE)
\begin{equation}
i \hbar \partial_t \psi_n(t)
=
H(t) \psi_n(t)
\label{tdse}
\end{equation}
to be solved using Floquet's solution \cite{shirley1965solution, sambe1973steady, rudner2020floquet}
\begin{equation}
\psi_n(t)
=
e^{-i \epsilon_n t / \hbar}
\Phi_n(t)
,
\Phi_n(t+T) = \Phi_n(t)
.
\label{floquet_ansatz}
\end{equation}
Substituting Eq.~\eqref{floquet_ansatz} into Eq.~\eqref{tdse} yields an eigenvalue problem
\begin{equation}
[H(t) - i \hbar \partial_t] \Phi_n(t) 
= 
\epsilon_n \Phi_n(t)
\label{floquet_eigen_time}
\end{equation}
for the quasienergies $\epsilon_n$, with the Floquet Hamiltonian given in brackets and the Floquet eigenstates denoted as $\Phi_n(t)$. By expanding time-periodic functions in Fourier series ($\Omega = 2 \pi / T$),
\begin{equation}
H(t)
=
\sum_{m}^{}
H^{(m)} e^{-i m \Omega t}
,
\label{ham_fourier}
\end{equation}
\begin{equation}
\Phi_n(t)
=
\sum_{m}^{}
\phi_n^{(m)} e^{-i m \Omega t}
,
\label{wavefunction_fourier}
\end{equation}
Eq.~\eqref{floquet_eigen_time} is transformed into a computationally tractable eigenvalue problem
\begin{equation}
\sum_{m'}^{}
\tilde{H}^{(m,m')}
\phi_n^{(m')}
=
\epsilon_n
\phi_n^{(m)}
\label{floquet_eigen_fourier}
\end{equation}
with the block elements given by
\begin{align}
\tilde{H}^{(m,m')}
&=
\frac{1}{T} 
\int_{0}^{T} dt 
H(t) e^{i (m-m') \Omega t}
-
\delta_{m m'} m \hbar \Omega \boldsymbol{1}
\notag
\\
&=
H^{(m-m')}
-
\delta_{m m'} m \hbar \Omega \boldsymbol{1}
.
\label{floquet_matrix_blocks}
\end{align}
$m$ and $m'$ take integral values from $-N_\textrm{Floquet}$ up to $N_\textrm{Floquet}$, where the truncation at $N_\textrm{Floquet} = 10$ ensures numerically converged quasienergy spectra for our system. Eqs.~\eqref{floquet_eigen_fourier} and~\eqref{floquet_matrix_blocks} can be written more compactly as
\begin{equation}
H^{(\textrm{Floquet})}
\boldsymbol{\phi}_n
=
\epsilon_n
\boldsymbol{\phi}_n
,
\label{floquet_eigen_fourier_compressed}
\end{equation}
where $H^{(\textrm{Floquet})}$ is the Floquet Hamiltonian in the Fourier space. As we will describe in detail later, under certain conditions a particular replica of the quasienergy spectrum can be interpreted as the Floquet-renormalized energy levels of the magnetic molecule.


\subsection{Assessment of Static Magnetic Field Stability}

We define an energy level as possessing static magnetic field stability (SMFS) if it exhibits the property
\begin{equation}
\left|
\frac{\partial \epsilon_n}{\partial \vec{B}_s}
\right|
=
0
\label{smfs_condition}
\end{equation}
for a particular $\vec{B}_s$. It has been demonstrated \cite{shiddiq2016enhancing, kundu2023electron, chen2025enhancing} that if the two energy levels forming a qubit have SMFS at the same $\vec{B}_s$ (clock transition), then spin decoherence is suppressed significantly \footnote{Mathematically, one only technically needs $| \partial (\epsilon_n-\epsilon_{n'}) / \partial \vec{B}_s | = 0$ ($n \neq n'$) to have a clock transition, which is a more relaxed condition. The optimization procedure in Appendix~\ref{sec:app-clock} could be generalized to check for this condition instead, and would potentially find additional solutions. However, we note that there are several physical reasons to impose the more restrictive definition, requiring each level to independently feature SMFS. If the energy gap is stable against magnetic noise, but the individual energy levels are not, then the corresponding quantum states are also varying randomly. This will create fluctuations in the decoherence mechanisms, including coupling of the electronic spin to the nuclear spin and phonon baths. Quantum computation will also be degraded since the Floquet qubit state initialization, single- and two-qubit gate operations, and state readout protocols are state dependent}. Computationally, the necessary energy gradients are obtained accurately and efficiently via a generalization of the Hellmann-Feynman theorem to Eq.~\eqref{floquet_eigen_fourier_compressed}
\begin{equation}
\frac{\partial \epsilon_n}{\partial (\vec{B}_s)_\alpha}
=
\boldsymbol{\phi}_n^\dagger
\left(
\frac{\partial H^{(\textrm{Floquet})}}{\partial (\vec{B}_s)_\alpha}
\right)
\boldsymbol{\phi}_n
,
\label{hf_theorem}
\end{equation}
where $\alpha$ refers separately to the $x$, $y$, and $z$ components. No finite-differences are required in this approach since
\begin{align}
\left(
\frac{\partial H^{(\textrm{Floquet})}}{\partial (\vec{B}_s)_\alpha}
\right)^{(m,m')}
&=
\frac{\partial \tilde{H}^{(m,m')}}{\partial (\vec{B}_s)_\alpha}
\notag
\\
&=
\mu_B
\left(
\tilde{g} \vec{s}
\right)_\alpha
\delta_{m m'}
.
\label{hf_matrix_blocks}
\end{align}


\subsection{Exact Numerical Effective Hamiltonian}

The exact effective Hamiltonian can be extracted numerically as follows \cite{bukov2015universal}. Consider a solution of the TDSE $\psi_n(t)$, see Eq.~\eqref{tdse}. The wave function is propagated from $t = 0$ to $t = T$ by making use of the one-cycle (Floquet) time-evolution operator
\begin{equation}
\psi_n(T)
=
U_F \psi_n(0)
=
U_F \Phi_n(0)
,
\label{one-cycle_evolution}
\end{equation}
\begin{equation}
U_F
\equiv
U(T,0)
=
\mathcal{T}
\textrm{exp}
\left\{
- \frac{i}{\hbar}
\int_{0}^{T} dt' H(t')
\right\}
,
\label{one-cycle_propagator}
\end{equation}
where $\mathcal{T}$ indicates time-ordering. At the same time, from Eq.~\eqref{floquet_ansatz} we have
\begin{equation}
\psi_n(T)
=
e^{-i \epsilon_n T / \hbar}
\Phi_n(T)
=
e^{-i \epsilon_n T / \hbar}
\Phi_n(0)
.
\label{one-cycle_floquet}
\end{equation}
Combining Eqs.~\eqref{one-cycle_evolution} and~\eqref{one-cycle_floquet} leads to an eigenvalue equation
\begin{equation}
U_F \Phi_n(0)
=
e^{-i \epsilon_n T / \hbar} \Phi_n(0)
.
\label{one-cycle_eigen}
\end{equation}
Now it is sensible to define the exact effective Hamiltonian via the relation
\begin{equation}
U_F
\equiv
\textrm{exp}
\left\{
- \frac{i}{\hbar}
T H^{\textrm{(eff)}}
\right\}
\label{one-cycle_propagator_eff}
\end{equation}
so that
\begin{equation}
H^{\textrm{(eff)}}
\equiv
\frac{i \hbar}{T}
\textrm{ln}(U_F)
\label{ham_eff_exact}
\end{equation}
and its eigenvalues will be $\epsilon_n$ \footnote{The natural logarithm of a complex valued matrix is computed numerically in Python using the convention of the scipy.linalg.logm function}. $U_F$ is evaluated numerically by splitting the time-evolution into $N_T = 100$ sufficiently small time steps $\Delta t$
\begin{equation}
U_F
\approx
\mathcal{T}
\prod_{N=1}^{N_T}
\textrm{exp}
\left\{
- \frac{i}{\hbar}
\Delta t H(t_N)
\right\}
,
\label{propagator_numerical}
\end{equation}
where $t_N = (N-1) \Delta t$ and $T = N_T \Delta t$.


\subsection{The Dynamical Cancellation Problem for $S = 1$}

Up to this point, the theory applies for general total spin $S$. For $S = 1$ specifically we have $2S+1 = 3$ so that the spin operators are $3 \times 3$ matrices each with nine elements. Explicitly, we have
\begin{equation}
s_x
=
\frac{1}{\sqrt{2}}
\begin{bmatrix}
0 & 1 & 0 \\
1 & 0 & 1 \\
0 & 1 & 0 \\
\end{bmatrix}
,
\label{spin1_x}
\end{equation}
\begin{equation}
s_y
=
\frac{1}{\sqrt{2}}
\begin{bmatrix}
0 & -i & 0 \\
i & 0 & -i \\
0 & i & 0 \\
\end{bmatrix}
,
\label{spin1_y}
\end{equation}
\begin{equation}
s_z
=
\begin{bmatrix}
1 & 0 & 0 \\
0 & 0 & 0 \\
0 & 0 & -1 \\
\end{bmatrix}
.
\label{spin1_z}
\end{equation}
Given the form of the static spin Hamiltonian (Eqs.~\eqref{ham_eq}-\eqref{ham_zeeman}), the effective Hamiltonian resulting from the Floquet driving should then be written as a linear combination of the following set of nine linearly independent basis matrices \footnote{Note that $\{ \cdots , \cdots \}$ indicates the anticommutator}
\begin{align}
\{ M_b \}
=
&\{
M_1, M_2, M_3, M_4, M_5, M_6, M_7, M_8, M_9
\}
\notag
\\
=
&\{
s_x^2
,
s_y^2
,
s_z^2
,
\{ s_x , s_y \}
,
\{ s_x , s_z \}
,
\{ s_y , s_z \}
,
\notag
\\
&\phantom{\{}
s_x
,
s_y
,
s_z
\}
.
\label{basis_matrices}
\end{align}
The effective Hamiltonian is expanded as
\begin{equation}
H^{\textrm{(eff)}}
=
\sum_{b=1}^{9}
c_b
M_b
.
\label{ham_eff_expansion}
\end{equation}
Together, Eqs.~\eqref{basis_matrices} and~\eqref{ham_eff_expansion} imply
\begin{equation}
H^{\textrm{(eff)}}
=
\vec{s}\,^T \tilde{D}^{\textrm{(eff)}} \vec{s}
+
\mu_B
\vec{\mathcal{B}}^{\textrm{(eff)}}
\cdot
\vec{s}
,
\label{ham_eff_spin1}
\end{equation}
where
\begin{equation}
\tilde{D}^{\textrm{(eff)}}
=
\begin{bmatrix}
c_1 & c_4 & c_5 \\
c_4 & c_2 & c_6 \\
c_5 & c_6 & c_3 \\
\end{bmatrix}
\label{zf_tensor_eff}
\end{equation}
and
\begin{equation}
\vec{\mathcal{B}}^{\textrm{(eff)}}
=
\frac{1}{\mu_B}
\left(
c_7 \hat{x}
+
c_8 \hat{y}
+
c_9 \hat{z}
\right)
.
\label{zeeman_eff}
\end{equation}
Physically, the Floquet drive can renormalize both the zero-field and the Zeeman terms. Once the exact effective Hamiltonian is constructed numerically using Eqs.~\eqref{ham_eff_exact} and ~\eqref{propagator_numerical}, we can devise a linear algebra problem to solve for the set of basis coefficients $\{ c_b \}$ (see Appendix~\ref{sec:app-exact}). 

We define the dynamical cancellation problem as finding the external static magnetic field $\vec{B}_s$ that leads to $\vec{\mathcal{B}}^{\textrm{(eff)}} = 0$. Equivalently, one needs to solve for $c_7 = c_8 = c_9 = 0$. We recognized the importance of solving this problem due to a general property of Hamiltonians of the form in Eq.~\eqref{ham_eff_spin1}, where $\tilde{D}^{\textrm{(eff)}}$ remains symmetric. By means of symbolic manipulations in Python, we showed analytically that the three energy levels have SMFS when $\vec{\mathcal{B}}^{\textrm{(eff)}} = 0$, provided that $\tilde{D}^{\textrm{(eff)}}$ is independent of $\vec{B}_s$ \footnote{For the proof we begin with Eqs.~\eqref{ham_eff_spin1}-\eqref{zeeman_eff} as well as the spin 1 matrices in Eqs.~\eqref{spin1_x}-\eqref{spin1_z} to write the Hamiltonian symbolically as a $3 \times 3$ matrix. Next, the three eigenvalues are derived in terms of all of the system parameters. Afterwards, the derivative of each of the three eigenvalues with respect to each of the three Cartesian components of $\vec{\mathcal{B}}^{\textrm{(eff)}}$ is taken. By substitution, all nine of these derivatives are zero at $\vec{\mathcal{B}}^{\textrm{(eff)}} = 0$. On the basis of the chain rule, if $\vec{\mathcal{B}}^{\textrm{(eff)}} = \vec{\mathcal{B}}^{\textrm{(eff)}}(\vec{B}_s)$ and $\tilde{D}^{\textrm{(eff)}}$ is independent of $\vec{B}_s$, then all three energy levels have SMFS}. We note that there could be other solutions. This would suggest that finding the levels with SMFS and solving the dynamical cancellation problem are equivalent for the Floquet system; however, later we will see that independently generating the numerical solution to both problems yields a slightly different result, implying a weak dependence of $\tilde{D}^{\textrm{(eff)}}$ on $\vec{B}_s$ for the system parameters under consideration.


\subsection{Approximate Analytical Effective Hamiltonian}

Some insights into the numerical results to follow from the previous subsections will be achieved by deriving an effective Hamiltonian analytically, which applies for sufficiently large Floquet photon energies $\hbar \Omega$. On the basis of second-order Van Vleck degenerate perturbation theory (see Appendix~\ref{sec:app-vV}), the effective Hamiltonian for general total spin $S$, a $g$-tensor $\tilde{g} = g \boldsymbol{1}$, and a time-periodic magnetic field of the form
\begin{equation}
\vec{B}(t)
=
\sum_{m}^{} \vec{B}^{(m)} e^{-i m \Omega t}
\label{magnetic_fourier_exp}
\end{equation}
is given by
\begin{equation}
H^{\textrm{(eff)}}
\approx
H_{\textrm{ZF}}^{\textrm{(eff)}} 
+ 
H_{\textrm{Zeeman}}^{\textrm{(eff)}}
+
H_{\textrm{neq}}
,
\label{ham_hf}
\end{equation}
\begin{equation}
H_{\textrm{ZF}}^{\textrm{(eff)}}
=
\vec{s}\,^T \tilde{D}^{\textrm{(eff)}} \vec{s}
,
\label{ham_zf_hf}
\end{equation}
\begin{equation}
H_{\textrm{ZF}}^{\textrm{(eff)}}
=
H_{\textrm{ZF}}
+
\Delta H_{\textrm{ZF}}
,
\label{ham_zf_breakdown_hf}
\end{equation}
\begin{equation}
\Delta H_{\textrm{ZF}}
=
\tilde{C}_1 \{ s_x , s_y \}
+
\tilde{C}_2 \{ s_x , s_z \}
+
\tilde{C}_3 \{ s_y , s_z \}
,
\label{ham_deltazf_hf}
\end{equation}
\begin{equation}
H_{\textrm{Zeeman}}^{\textrm{(eff)}}
=
\mu_B g \vec{B}^{\textrm{(eff)}} \cdot \vec{s}
,
\label{ham_zeeman_hf}
\end{equation}
\begin{align}
H_{\textrm{neq}}
=
\tilde{C}_4 [ \{ s_y , s_z \} , s_x ]
&+
\tilde{C}_5 [ \{ s_x , s_z \} , s_y ]
\notag
\\
&+
\tilde{C}_6 [ \{ s_x , s_y \} , s_z ]
.
\label{ham_neq_hf}
\end{align}
The last equation is labeled as non-equilibrium (neq) because it may not be reducible to a form that merely renormalizes the terms already present in the static Hamiltonian. We elucidate the general mathematical distinction between $\vec{\mathcal{B}}^{\textrm{(eff)}}$ from the previous subsection and $\vec{B}^{\textrm{(eff)}}$ in Appendix~\ref{sec:app-cancel} for an arbitrary $g$-tensor $\tilde{g}$. The dependence of the coefficients appearing in Eqs.~\eqref{ham_deltazf_hf} and~\eqref{ham_neq_hf} on the static spin and Floquet drive parameters is given as follows
\begin{equation}
\tilde{C}_\ell
\equiv
\frac{(\mu_B g)^2}{2 (\hbar \Omega)^2}
\sum_{m \neq 0}^{}
\frac{1}{m^2}
C_{\ell m}
,
\label{coeff_ell}
\end{equation}
\begin{align}
C_{1 m}
\equiv
- (D+E) &B_x^{(-m)} B_y^{(m)}
\notag
\\
- (D-E) &B_y^{(-m)} B_x^{(m)}
,
\label{coeff_1m}
\end{align}
\begin{align}
C_{2 m}
\equiv
(D+E) &B_x^{(-m)} B_z^{(m)}
\notag
\\
+ 2E &B_z^{(-m)} B_x^{(m)}
,
\label{coeff_2m}
\end{align}
\begin{align}
C_{3 m}
\equiv
(D-E) &B_y^{(-m)} B_z^{(m)}
\notag
\\
-2E &B_z^{(-m)} B_y^{(m)}
,
\label{coeff_3m}
\end{align}
\begin{equation}
C_{4 m}
\equiv
-i (D+E) B_x^{(-m)} B_x^{(m)}
,
\label{coeff_4m}
\end{equation}
\begin{equation}
C_{5 m}
\equiv
i (D-E) B_y^{(-m)} B_y^{(m)}
,
\label{coeff_5m}
\end{equation}
\begin{equation}
C_{6 m}
\equiv
2i E B_z^{(-m)} B_z^{(m)}
.
\label{coeff_6m}
\end{equation}
Furthermore, the effective magnetic field has contributions from all terms in the Van Vleck expansion
\begin{equation}
\vec{B}^{\textrm{(eff)}}
=
\vec{B}_s
+
\vec{B}^{(0)}
+
\vec{B}^{\textrm{(eff)}}_{1}
+
\vec{B}^{\textrm{(eff)}}_{2,1}
+
\vec{B}^{\textrm{(eff)}}_{2,2}
,
\label{magnetic_eff_hf}
\end{equation}
which are given as
\begin{equation}
\vec{B}^{\textrm{(eff)}}_{1}
=
\frac{i \mu_B g}{2 \hbar \Omega}
\sum_{m \neq 0}^{}
\frac{1}{m}
\left(
\vec{B}^{(-m)}
\times
\vec{B}^{(m)}
\right)
,
\label{magnetic_eff_1_hf}
\end{equation}
\begin{align}
\vec{B}^{\textrm{(eff)}}_{2,1}
=
&\frac{(\mu_B g)^2}{2 (\hbar \Omega)^2}
\sum_{m \neq 0}^{} 
\frac{1}{m^2}
\notag
\\
&\left[
\vec{B}^{(m)}
\times
\left(
\vec{B}^{(-m)}
\times
\vec{B}_s
\right)
\right]
,
\label{magnetic_eff_21_hf}
\end{align}
\begin{align}
\vec{B}^{\textrm{(eff)}}_{2,2}
=
&\frac{(\mu_B g)^2}{3 (\hbar \Omega)^2}
\sum_{m \neq 0}^{}
\sum_{n \neq 0, m}^{}
\frac{1}{mn}
\notag
\\
&\left[
\vec{B}^{(n)}
\times
\left(
\vec{B}^{(-m)}
\times
\vec{B}^{(m-n)}
\right)
\right]
.
\label{magnetic_eff_22_hf}
\end{align}
So far in this subsection the results apply for general total spin $S$. In particular, if $S = 1$ specifically we have a simplification of the non-equilibrium term (Eq.~\eqref{ham_neq_hf}) to
\begin{align}
H_{\textrm{neq}}
=
2i (\tilde{C}_6 - \tilde{C}_5) s_x^2
&+
2i (\tilde{C}_4 - \tilde{C}_6) s_y^2
\notag
\\
&+
2i (\tilde{C}_5 - \tilde{C}_4) s_z^2
.
\label{ham_neq_spin1_hf}
\end{align}
Importantly, later we will utilize the results from this subsection to emphasize that the primary effect of the Floquet drive is not to introduce an effective Zeeman shift, which could otherwise simply be produced by applying an external static magnetic field.


\section{Results and Analysis}
\label{sec:results}

We will illustrate how Floquet engineering modifies the energy levels of a magnetic molecule using a set of simulation parameters that can be experimentally realized with existing molecules \cite{bayliss2020optically}. Some examples of $S = 1$ systems where clock transitions have been explicitly demonstrated experimentally include $\textrm{Cr}_7 \textrm{Mn}$ \cite{collett2019clock} and Ni(II) \cite{rubin2021chemical} magnetic molecules.

A summary of the spin and Floquet control parameters chosen for this work is conveniently summarized in Table~\ref{tab:parameters}. For the static magnetic molecule we consider an emergent total electronic spin of $S = 1$ with $D =$ 5 $\mu$eV and $E/D =$ 0, 0.1, 1/3 for the zero-field splitting parameters, and $\tilde{g} = 2 * \boldsymbol{1}$ for the $g$-tensor \footnote{One can trivially see the results for $D < 0$ by multiplying the time-dependent Hamiltonian through by -1}. The choice of $S = 1$ is important because the pure electronic spin Hamiltonian encompassed by Eqs.~\eqref{ham_eq}-\eqref{ham_zeeman} then features three energy levels with SMFS (see Eq.~\eqref{smfs_condition}) at $\vec{B}_s = 0$, where a pair of them would function as a clock transition qubit. This was also implied from the discussion in Sec.~\ref{sec:theory}E. While $S = 1/2$ would have exactly the two energy levels needed to form a qubit, they do not display SMFS. In this sense, selecting $S = 1$ produces a minimal model for studying SMFS. Overall, as discussed before, constructing a qubit from two energy levels that have SMFS leads to a significant suppression of spin decoherence \cite{shiddiq2016enhancing, kundu2023electron, chen2025enhancing}.

Given the time-periodicity condition in Eq.~\eqref{time-periodic}, in the most general case the dynamical external magnetic field is a Fourier series given by Eq.~\eqref{magnetic_fourier_exp}. Equivalently,
\begin{align}
\vec{B}(t)
=
\vec{B}^{(0)}
+
\sum_{m=1}^{\infty}
\Big[
&\vec{B}_\textrm{cos}^{(m)} \, \textrm{cos}(m \Omega t)
\notag
\\
&+
\vec{B}_\textrm{sin}^{(m)} \, \textrm{sin}(m \Omega t)
\Big]
,
\label{magnetic_fourier_cos-sin}
\end{align}
with the conversion provided by ($m \neq 0$)
\begin{equation}
\vec{B}^{(m)}
=
\frac{
\vec{B}_\textrm{cos}^{(|m|)}
+
i
\,
\textrm{sgn}(m)
\vec{B}_\textrm{sin}^{(|m|)}
}{2}
.
\label{magnetic_fourier_conversion}
\end{equation}
We consider a monochromatic source, and improve the above notation by separating the amplitude and polarization of the Floquet drive
\begin{equation}
\vec{B}(t)
=
B_F
\left(
\vec{P}^{(-1)} e^{i \Omega t}
+
\vec{P}^{(1)} e^{-i \Omega t}
\right)
,
\label{magnetic_mono_exp}
\end{equation}
\begin{equation}
\vec{B}(t)
=
B_F
\left[
\vec{P}_\textrm{cos} \, \textrm{cos}(\Omega t)
+
\vec{P}_\textrm{sin} \, \textrm{sin}(\Omega t)
\right]
,
\label{magnetic_mono_cos-sin}
\end{equation}
\begin{equation}
\vec{P}^{(\pm 1)}
=
\frac{
\vec{P}_\textrm{cos}
\pm
i
\,
\vec{P}_\textrm{sin}
}{2}
.
\label{polarization_vector_conversion}
\end{equation}
$B_F$ is the Floquet amplitude and the vectors $\vec{P}^{(-1)}$, $\vec{P}^{(1)}$, $\vec{P}_\textrm{cos}$, and $\vec{P}_\textrm{sin}$ are the Floquet polarization vectors. Depending on the polarization state, the Floquet amplitude may range from zero up to 300 mT. A number of polarizations are tested, including linear polarizations along the principal axes ($x$, $y$, $z$), linear polarizations at $45^\circ$ between principal axes ($+x+y$, $+x-y$, $+x+z$, $+x-z$, $+y+z$, $+y-z$), and circular polarizations in the planes relative to the principal axes ($(xy)+$, $(xy)-$, $(xz)+$, $(xz)-$, $(yz)+$, $(yz)-$). In the latter grouping, the sign indicates the handedness of the circular polarization. For example, $(xy)+$ means that the magnetic field vector rotates counter-clockwise as viewed from the $+z$ direction. Lastly, the Floquet photon energy $\hbar \Omega$ is taken to be 20 $\mu$eV, which is considerably larger than the total energy extent of the three levels from the static case ($D+E$). Under those conditions, the \textit{unfolded quasienergies} satisfying the relation
\begin{equation}
\lim_{B_F \to 0} 
\epsilon_n
=
E_n
\label{unfolded}
\end{equation}
can be physically interpreted as the Floquet-renormalized energy levels, see Eqs.~\eqref{tise} and~\eqref{floquet_eigen_time}. That is, when the Floquet problem is solved with the driving amplitude limiting to zero, the quasienergy replica that gives back the static energy eigenvalue is the physically relevant level. In practice, the Floquet amplitude is slowly incremented from zero so that the physically meaningful energy levels are tracked correctly. A similar approach was taken in previous work \cite{cupo2021floquet, cupo2023optical}, and a complete discussion of this subtlety in Floquet theory is provided in Sec. III of \cite{cupo2023optical}.


\subsection{Floquet Renormalized Clock Transitions}

\begin{figure*}[!htb]
\centering{\includegraphics[scale=0.45]{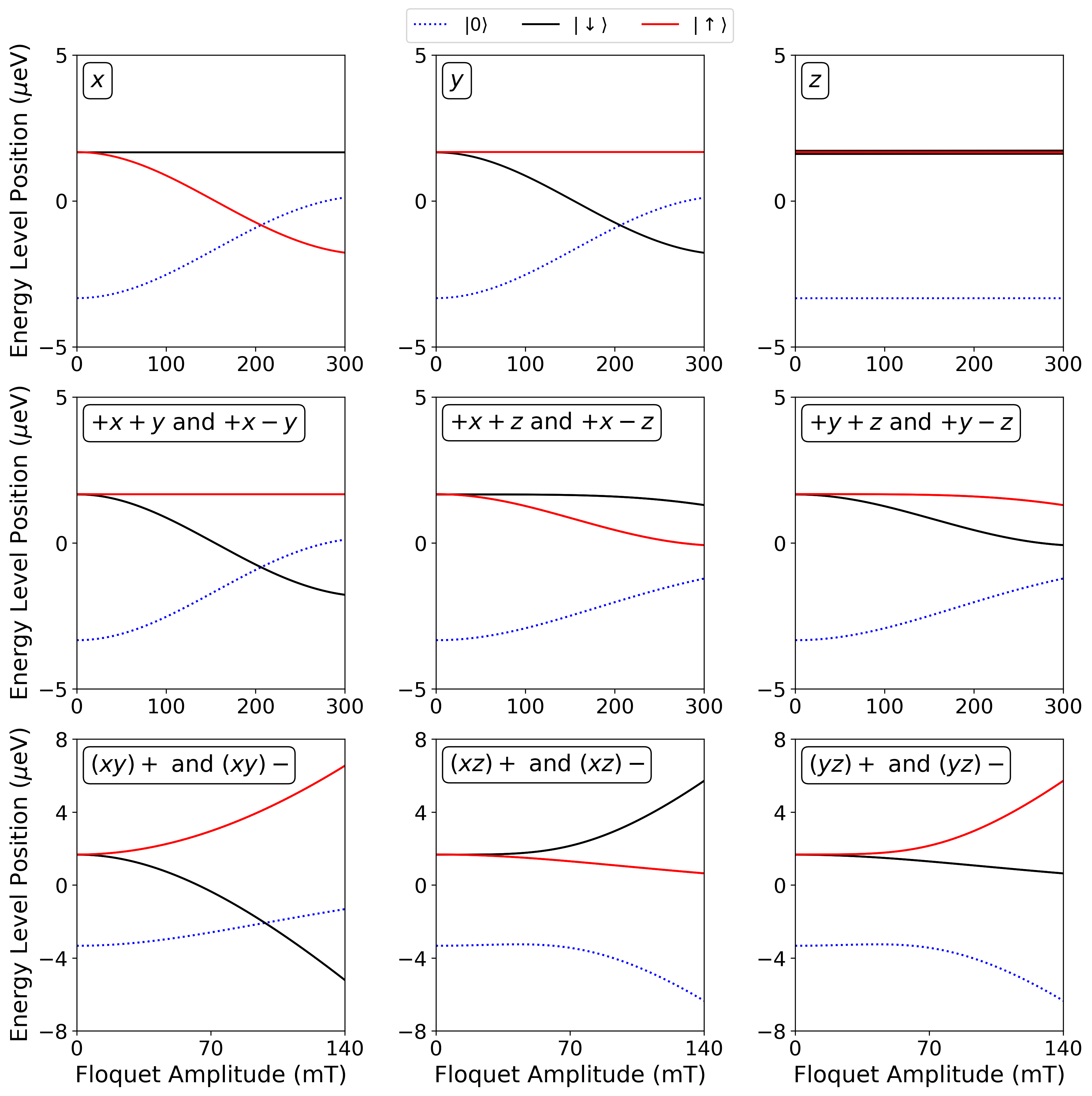}}
\caption{{\bf Renormalization of the three energy levels of an $S = 1$ magnetic molecule by Floquet driving.} The zero-field splitting parameters are $D =$ 5 $\mu$eV and $E/D =$ 0, and the external static magnetic field is zero. The top left inset of each panel indicates the polarization of the Floquet drive, with linear and circular polarizations in the top six and bottom three plots, respectively. Specific details about the polarizations are provided between Eqs.~\eqref{polarization_vector_conversion} and~\eqref{unfolded} in the main text. The labeling of each of the energy curves is explained in Appendix~\ref{sec:app-state}. The essential finding of our work is that for all linear polarizations and amplitudes of the Floquet control, the three energy levels feature static magnetic field stability (SMFS) as defined by Eq.~\eqref{smfs_condition}, which is crucial for suppressing spin decoherence.} 
\label{fig:energy_levels_ED0}
\end{figure*}

\begin{figure*}[!htb]
\centering{\includegraphics[scale=0.45]{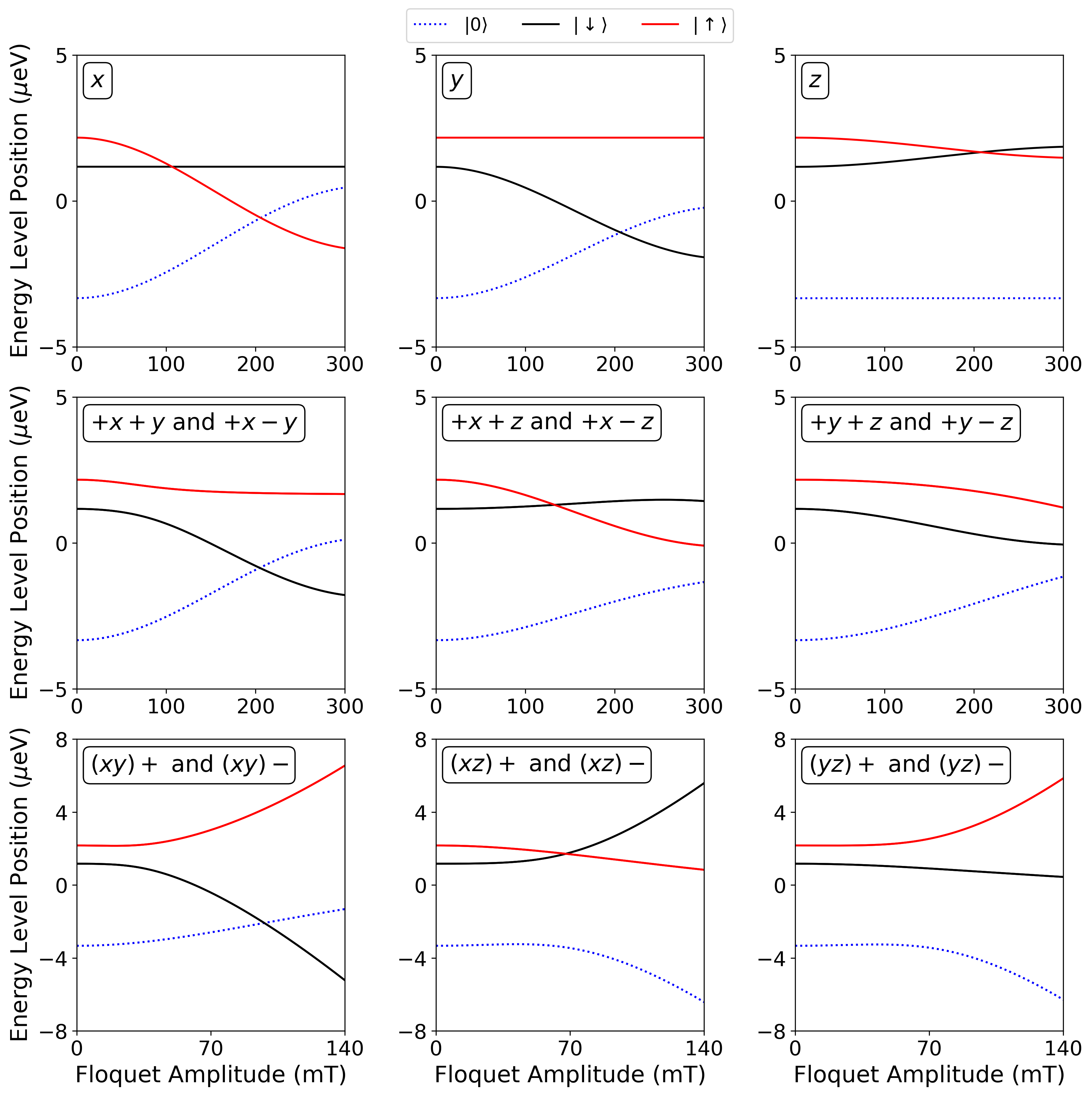}}
\caption{{\bf Renormalization of the three energy levels of an $S = 1$ magnetic molecule by Floquet driving.} Same as Fig.~\ref{fig:energy_levels_ED0} except $E/D =$ 0.1.} 
\label{fig:energy_levels_ED0.1}
\end{figure*}

\begin{figure*}[!htb]
\centering{\includegraphics[scale=0.45]{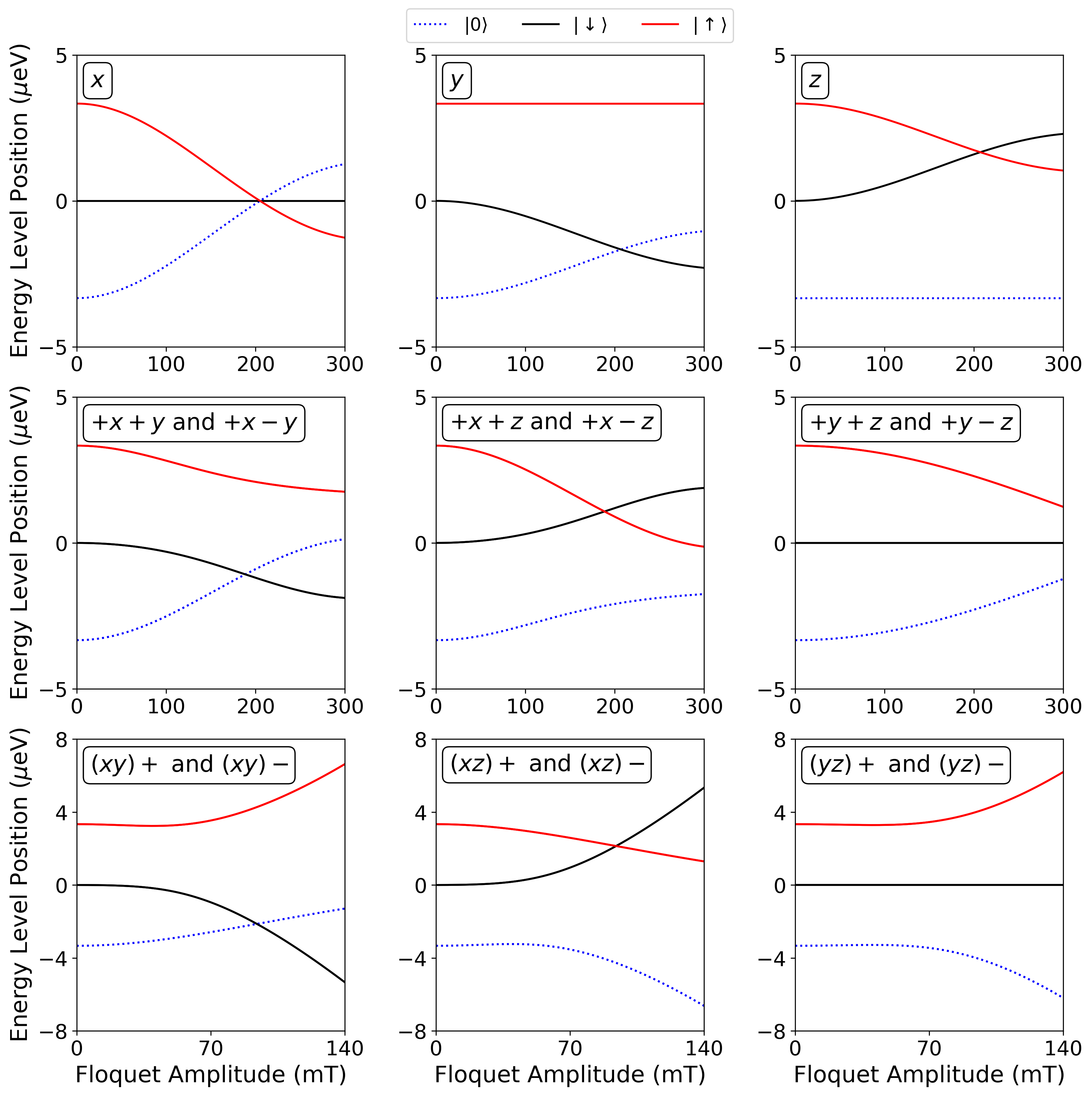}}
\caption{{\bf Renormalization of the three energy levels of an $S = 1$ magnetic molecule by Floquet driving.} Same as Fig.~\ref{fig:energy_levels_ED0} except $E/D =$ 1/3.} 
\label{fig:energy_levels_EDone-third}
\end{figure*}

In Figs.~\ref{fig:energy_levels_ED0}–\ref{fig:energy_levels_EDone-third}, we present the renormalized energy levels as a function of the Floquet amplitude for different drive polarizations, all with zero external static magnetic field, corresponding to the ratios $E/D = 0$, 0.1, and 1/3, respectively. An explanation for how the energy curves are labeled by state is included as Appendix~\ref{sec:app-state}. For each of the $E/D$ ratios we see that the differences between energy levels can be tuned continuously anywhere from zero up to $D+E$ for the linear polarizations and from zero up to 12 $\mu$eV for the circular polarizations. Practically, if an electromagnetic source is only available at a specific frequency in the GHz regime, the energy difference between two levels can be tuned to enable resonant addressing of a defined qubit. Perhaps the most useful case is applying $z$ polarization with $E/D = 0.1$ or 1/3, since the bottom energy level is unchanged and the separation between the two top levels varies from $2E$ down to zero \cite{bayliss2020optically}.

Additionally, when the Floquet drive is polarized along one of the principal axes, at least one of the three energy levels is unaffected by the external control. In the most extreme case, when $E/D = 0$ and the polarization is along the $z$ axis, all three levels are unchanged. Later we rationalize this property on the basis of the derived analytical effective Hamiltonian. For the $+y \pm z$ and $(yz)\pm$ polarizations, the middle energy level has this property but only for $E/D = 1/3$. Furthermore, with the exception of the $z$ polarization, the Floquet drive lifts the degeneracy between the upper two levels when $E/D = 0$. For $E/D = 1/3$, the $x$ polarization engineers a triple degeneracy point at around a Floquet amplitude of 200 mT. Moreover, in many cases the Floquet drive facilitates a crossing between two energy levels. In practice, when the control amplitude is modulated from zero up through a critical value where the crossing occurs, there will be some systematic transfer of occupation between the energy levels, in analogy to the Landau-Zener problem \cite{wittig2005landau, vitanov1999transition, ge2021universal}.

For each $E/D$ ratio, polarization, and Floquet amplitude combination, we check the condition in Eq.~\eqref{smfs_condition} by computing the relevant gradient for each energy level to determine potential SMFS. We find numerically that when the Floquet drive is linearly polarized, all three energy levels conveniently universally display SMFS at $\vec{B}_s = 0$. For $E/D = 0$ this is particularly useful because the Floquet drive separates the upper two energy levels, which then enables a clock transition qubit to be formed that was non-existent in the static case due to the level degeneracy (Fig.~\ref{fig:energy_levels_ED0}).

\begin{figure*}[!htb]
\centering{\includegraphics[scale=0.45]{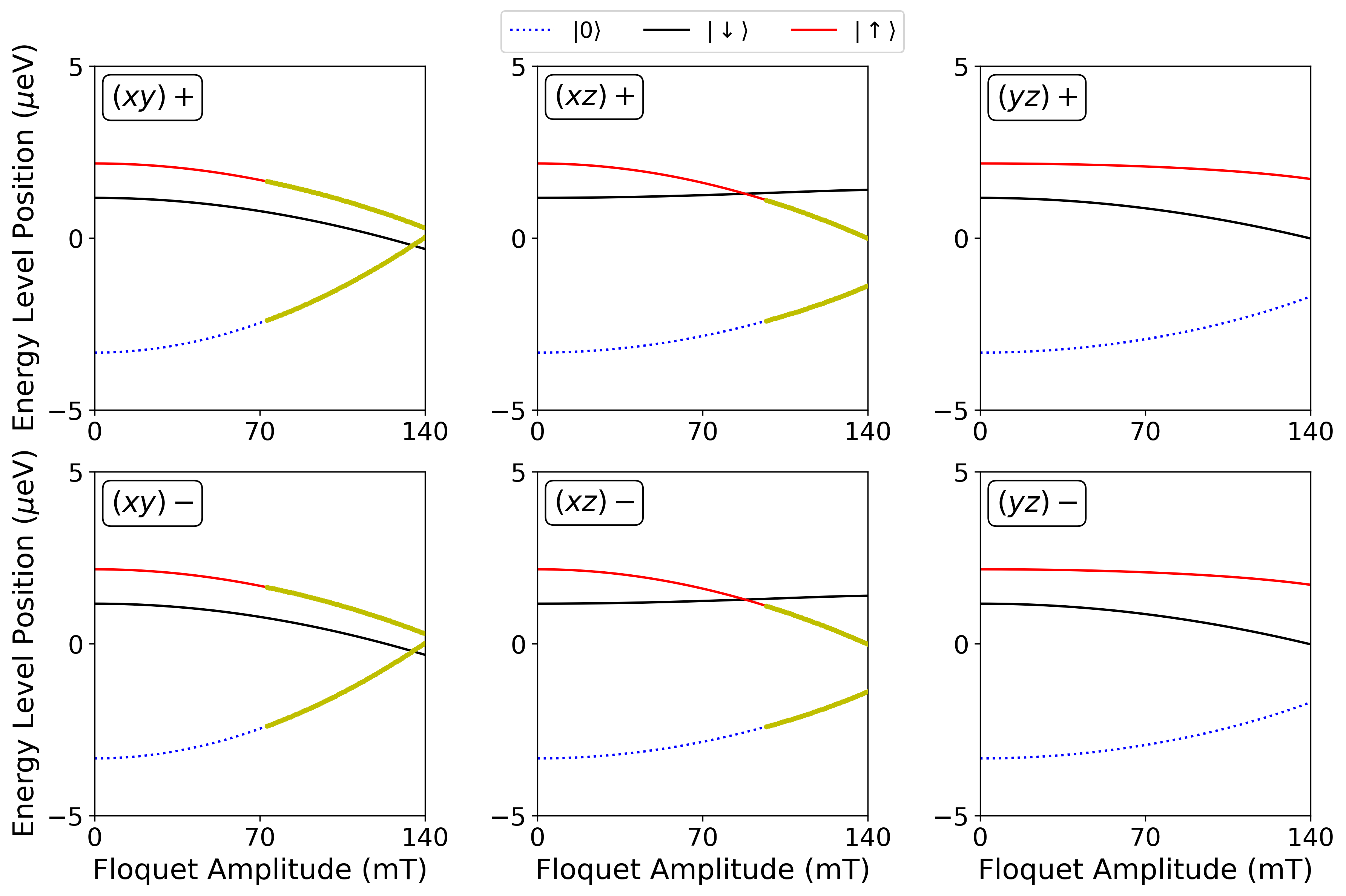}}
\caption{{\bf Renormalization of energy levels for circularly polarized Floquet control with possible SMFS.} Taking the parameters from Fig.~\ref{fig:energy_levels_ED0.1} ($E/D =$ 0.1), the external static magnetic field vector components (see next figure) are varied until all three energy levels are as close as possible to displaying SMFS. If the gradient magnitude from Eq.~\eqref{smfs_condition} is less than $10^{-2} \mu \textrm{eV} / \textrm{mT}$ for a given energy level, then we assign SMFS in this figure. If this condition is violated, the points on the energy curves are marked in yellow to indicate that there is no SMFS. Note that for the linear polarizations in Figs.~\ref{fig:energy_levels_ED0}-\ref{fig:energy_levels_EDone-third}, the computed gradient magnitudes in Eq.~\eqref{smfs_condition} are all less than $10^{-9} \mu \textrm{eV} / \textrm{mT}$.} 
\label{fig:clock_circular_energy}
\end{figure*}

\begin{figure*}[!htb]
\centering{\includegraphics[scale=0.45]{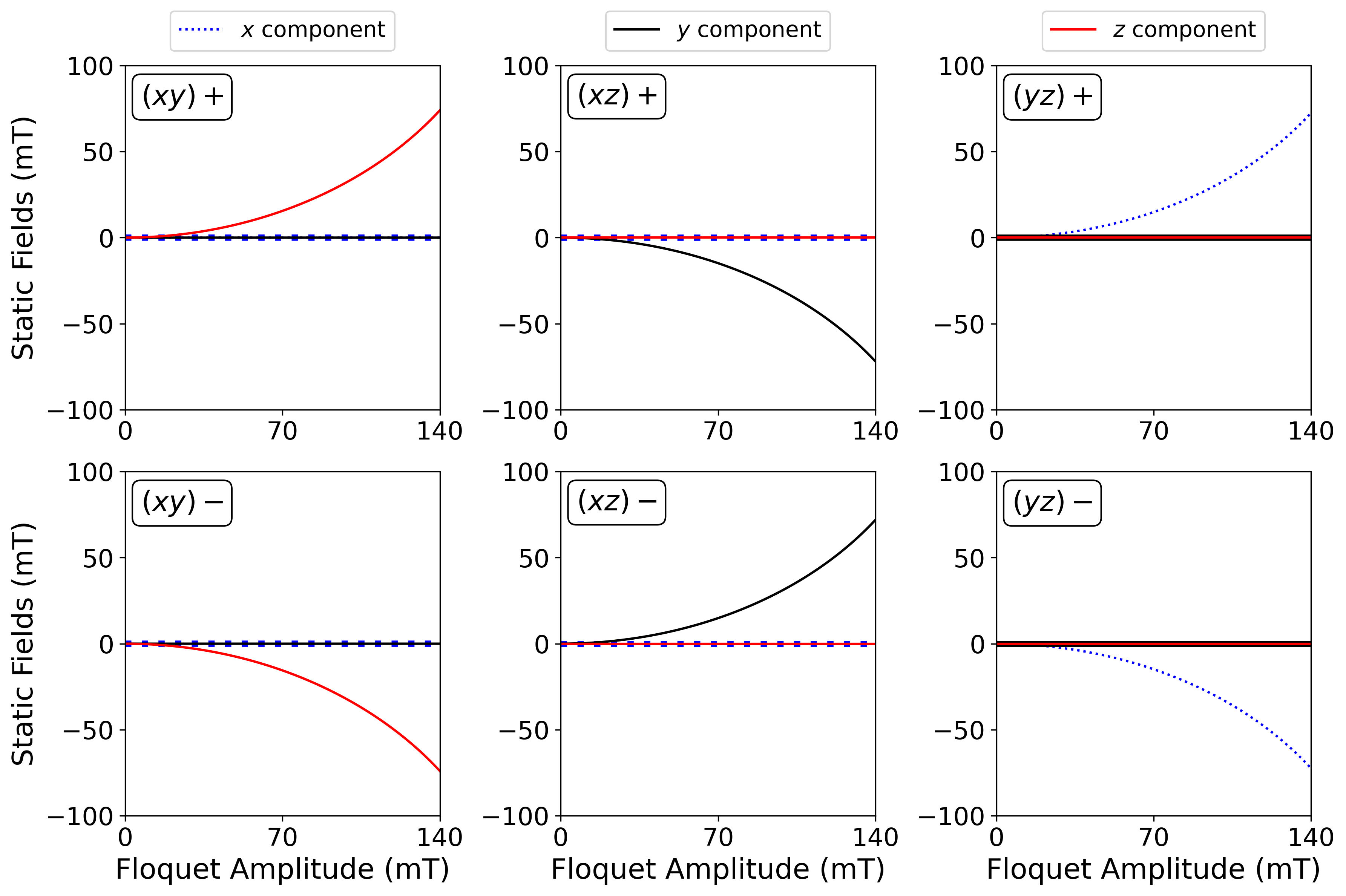}}
\caption{{\bf Corresponding external static magnetic field vector components required to realize the energy curves in the previous figure.} Only one component will be non-zero and is along the axis perpendicular to the plane of the circulation. The direction is reversed between the different helicities.} 
\label{fig:clock_circular_fields}
\end{figure*}

On the other hand, for circular polarization the three energy levels do not have SMFS at $\vec{B}_s = 0$, except perhaps for control amplitudes very close to zero. By means of the approach described in Appendix~\ref{sec:app-clock}, a numerical search was performed to find non-zero $\vec{B}_s$ that produce SMFS for the circularly polarized Floquet drive. The results for the energy levels and the corresponding vector components of the external static magnetic field required to produce them are provided in Figs.~\ref{fig:clock_circular_energy} and~\ref{fig:clock_circular_fields}, respectively for $E/D = 0.1$. It is apparent that the handedness does not affect the energy level scaling with the Floquet amplitude. Whichever plane the magnetic field vector rotates in, the required static magnetic field will be purely along the orthogonal principal axis, with opposing directions for the different handedness. The $(xy)\pm$ and $(yz)\pm$ polarizations are not particularly interesting since the lower level moves upwards, while the two upper levels shift downwards nearly parallel. However, for $(xz)\pm$ the spacing between the two upper states can be tuned from $2E$ to zero \cite{bayliss2020optically}. 

\begin{figure*}[!htb]
\centering{\includegraphics[scale=0.45]{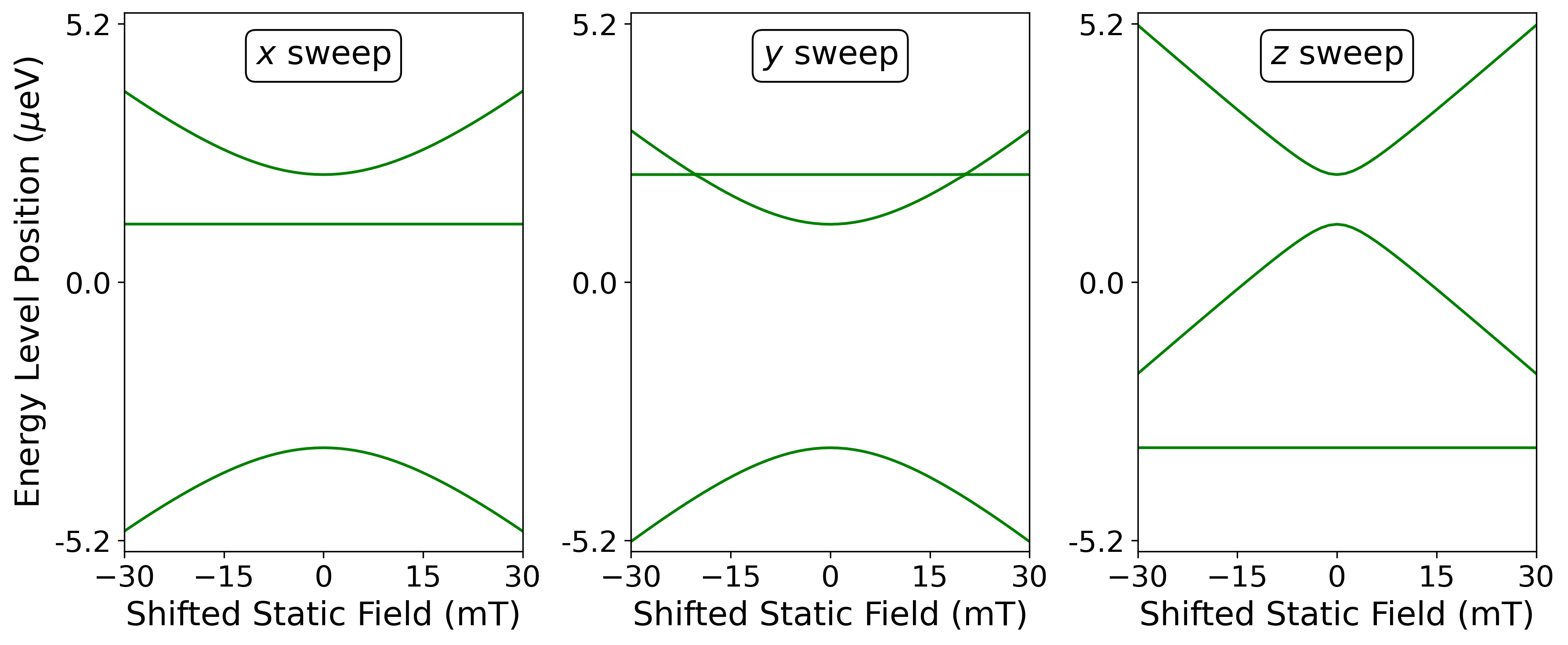}}
\caption{{\bf Energy sweep in the external static magnetic field for the non-driven system for $E/D =$ 0.1.} For a fixed Floquet amplitude of zero (static limit), the three energy levels are plotted as a function of the external static magnetic field, with the optimized vector components along the $x$, $y$, and $z$ principal axes shifted to zero. For the static case, the zero of the horizontal axes corresponds to zero external static magnetic field (so no actual shift was required). Along each direction the slope of all energy curves is zero at the zero of the horizontal axes, indicating that all three energy levels feature SMFS defined by Eq.~\eqref{smfs_condition}.} 
\label{fig:energy_sweep_eq}
\end{figure*}

\begin{figure*}[!htb]
\centering{\includegraphics[scale=0.45]{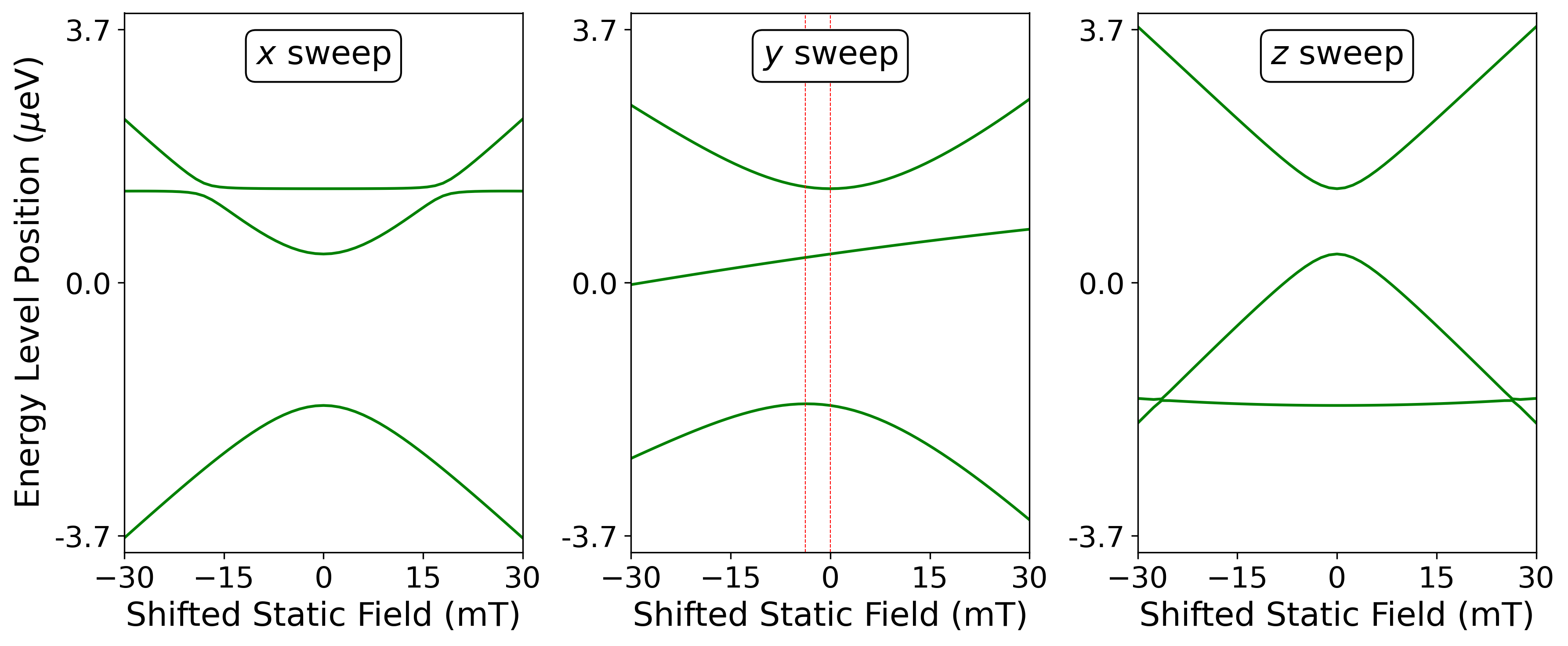}}
\caption{{\bf Energy sweep in the external static magnetic field for the $(xz)+$ polarization and a Floquet amplitude of 125 mT for $E/D =$ 0.1.} The three energy levels are plotted as a function of the external static magnetic field, with the optimized vector components along the $x$, $y$, and $z$ principal axes shifted to zero. While the upper level features SMFS, the other two do not at the zero of the horizontal axes. Along the $y$ direction, the relative maximum of the lower level is shifted to -3.8 mT, and the middle level is completely tilted.} 
\label{fig:energy_sweep_floquet}
\end{figure*}

There is a major caveat to emphasize here: In Fig.~\ref{fig:clock_circular_energy} any points marked in yellow have a gradient magnitude (see Eq.~\eqref{smfs_condition}) that is greater than $10^{-2} \mu \textrm{eV} / \textrm{mT}$, which we label as not having SMFS. To gain more insight, for a fixed Floquet polarization and amplitude we plot the energy levels as a function of the external static magnetic field around the optimized value. Fig.~\ref{fig:energy_sweep_eq} contains the reference result for zero driving amplitude, which applies for all polarizations. We see as expected that the slope of all three curves in each of the Cartesian directions at the optimized external static magnetic field is zero. In Fig.~\ref{fig:energy_sweep_floquet} we present the same plots but for the $(xz)+$ polarization with a Floquet amplitude of 125 mT. Not surprisingly, because of time-reversal symmetry breaking, the issue arises in the $y$ sweep: While the upper level has SMFS at the optimization point (zero on the horizontal axis), the lower level does not because the relative maximum occurs at -3.8 mT instead of zero as indicated by the vertical dashed lines, and the middle level is completely tilted. By inspecting the same plots for other polarizations and amplitudes, we find that this kind of behavior explains the presence of the yellow points in general. Importantly, if the gradient magnitude cutoff is raised to $10^{-3} \mu \textrm{eV} / \textrm{mT}$, the circularly polarized Floquet control can practically only produce one level with SMFS. Then a low spin decoherence qubit based on two levels each with SMFS would not be possible to engineer in that case. For comparison, with the linearly polarized control, the largest gradient magnitude resulting across all calculations is $< 10^{-9} \mu \textrm{eV} / \textrm{mT}$. 

To give physical context to the values of the energy gradients above, we refer to Shiddiq \textit{et al}. \cite{shiddiq2016enhancing}. While they study a specific molecular system (HoW$_{10}$), their results at least provide us with order of magnitude estimates. In particular, they find that if one shifts the $z$ component of the external static magnetic field 2 mT from a clock transition point, the measured $T_2$ time is degraded by more than a factor of two (see their Extended Data Figure 1b). On the basis of their simple energy model, see their Eq. 2, we calculate that the derivative of the clock transition energy gap with respect to the $z$ component of the external static magnetic field at this point is about 0.02 $\mu \textrm{eV} / \textrm{mT}$. For the $T_2$ time to not be degraded from the value at the clock transition, the maximal shift appears to be about 0.2 mT, which would put the derivative at about 0.002 $\mu \textrm{eV} / \textrm{mT}$. Therefore the cutoff values we are using ($10^{-2} \mu \textrm{eV} / \textrm{mT}$ and $10^{-3} \mu \textrm{eV} / \textrm{mT}$) are in fact near the boundary of defining a clock transition practically. We note that the largest values emerging for the linearly polarized control ($10^{-9} \mu \textrm{eV} / \textrm{mT}$) are many orders of magnitude smaller than these values. While it is difficult to say if we have just reached the numerical precision limit in our simulations, practically the values are effectively zero by comparison.


\subsection{Comparison to the Dynamical Cancellation Problem}

\begin{figure*}[!htb]
\centering{\includegraphics[scale=0.45]{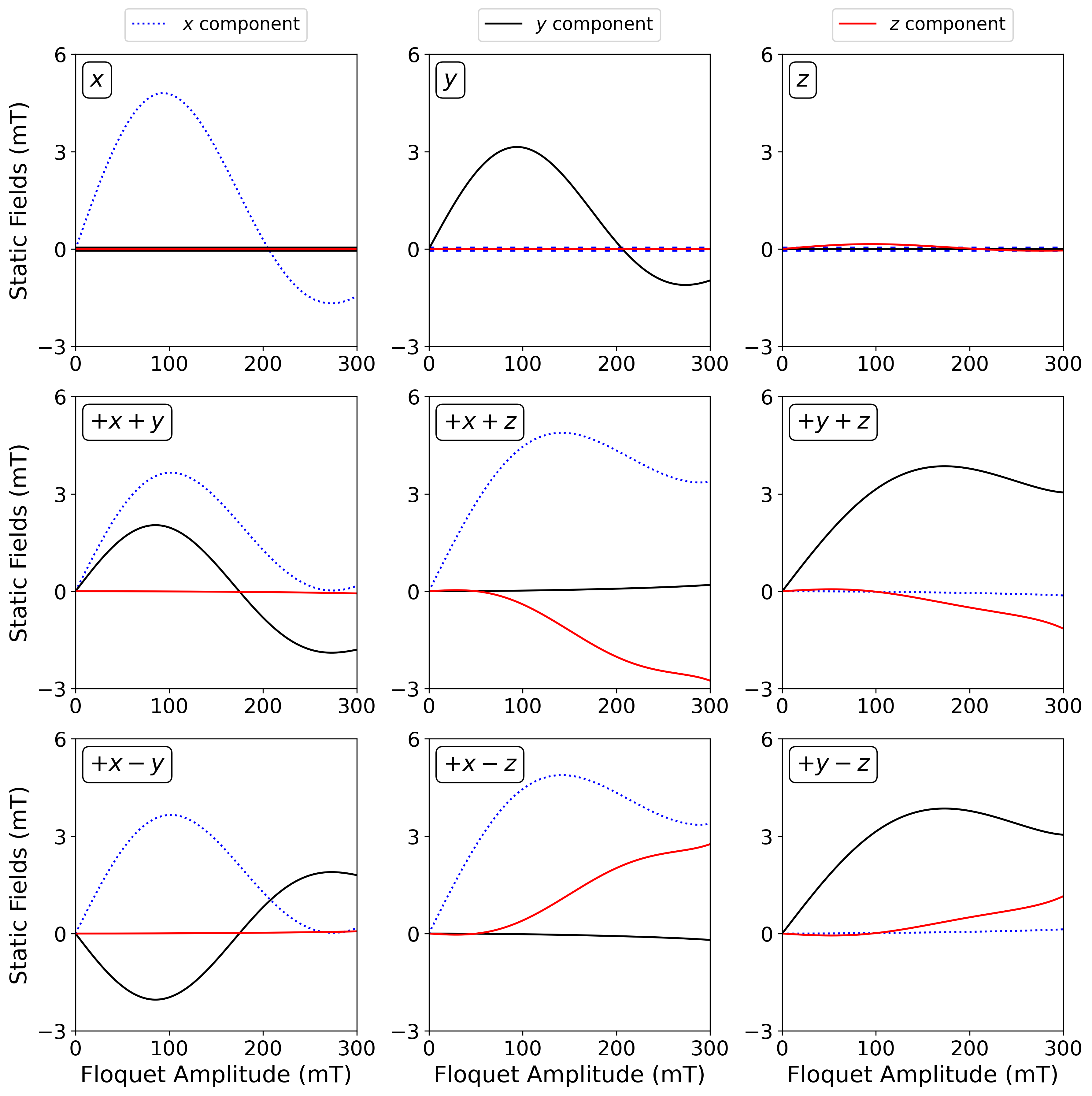}}
\caption{{\bf Vector components of the external static magnetic field required to solve the dynamical cancellation problem defined in Sec.~\ref{sec:theory}E of the main text for linear polarizations of the Floquet drive.} The components are all at most a few mT. On the other hand, for the levels to have SMFS, all of the components would be zero. Therefore, the two problems are closely related, but not the same, which can be explained by the effective zero-field tensor acquiring a weak dependence on the external static magnetic field.} 
\label{fig:cancel_linear}
\end{figure*}

\begin{figure*}[!htb]
\centering{\includegraphics[scale=0.45]{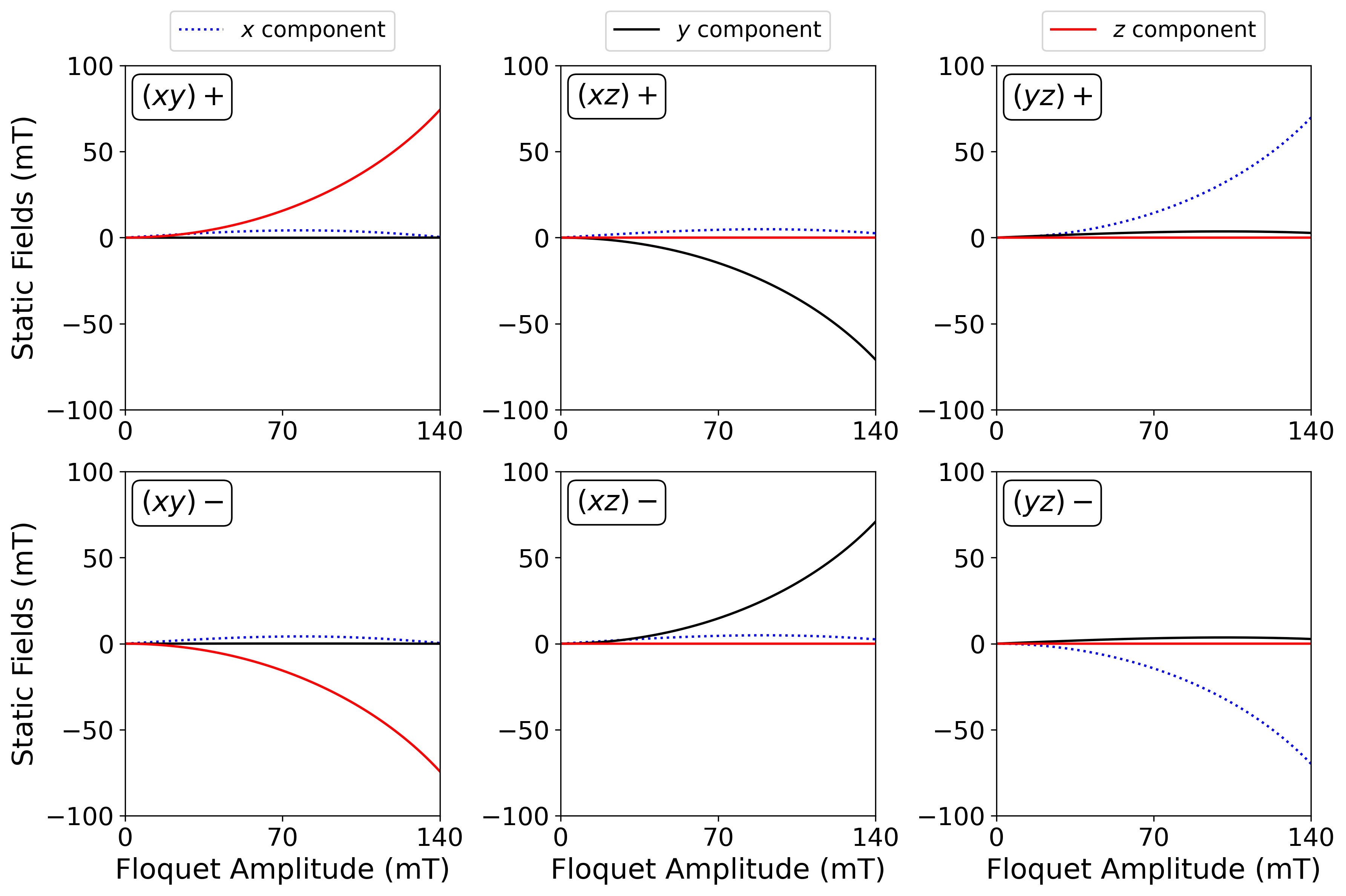}}
\caption{{\bf Vector components of the external static magnetic field required to solve the dynamical cancellation problem defined in Sec.~\ref{sec:theory}E of the main text for circular polarizations of the Floquet drive.} The components are very similar to the solutions for the SMFS problem provided in Fig.~\ref{fig:clock_circular_fields}. Therefore, the two problems are closely related, but not the same, which can be explained by the effective zero-field tensor acquiring a weak dependence on the external static magnetic field.} 
\label{fig:cancel_circular}
\end{figure*}

Using the iterative self-consistent approach described in Appendix~\ref{sec:app-cancel}, we solve the dynamical cancellation problem defined in Sec.~\ref{sec:theory}E for the three $E/D$ ratios with all combinations of the Floquet polarization and amplitude. At the optimization point, the energy levels as a function of the Floquet amplitude are nearly the same as those found from the SMFS problem and are therefore not worth visualizing. However, it is instructive to analyze the corresponding components of the external static magnetic field required to solve the dynamical cancellation problem. As an example, for $E/D = 0.1$ the solutions for the linear and circular polarizations are provided in Figs.~\ref{fig:cancel_linear} and~\ref{fig:cancel_circular}, respectively. For linear polarization, the levels with SMFS all occur at $\vec{B}_s = 0$, and for the dynamical cancellation problem all of the Cartesian components of $\vec{B}_s$ are only a few mT at the most (Fig.~\ref{fig:cancel_linear}). Similarly for circular polarization, a small difference in the solutions between the two problems can be seen by comparing Figs.~\ref{fig:clock_circular_fields} and~\ref{fig:cancel_circular}. Therefore, the SMFS and the dynamical cancellation problems are almost the same, at least for the range of parameters explored in this work. As discussed previously in Sec.~\ref{sec:theory}E, this observation implies that the Floquet drive gives $\tilde{D}^{\textrm{(eff)}}$ a weak dependence on $\vec{B}_s$.


\subsection{Clock Transition Gap Modification via Floquet Renormalization of the Zero-Field Tensor}

An essential question to answer is: Does the Floquet drive simply induce an effective Zeeman shift, in which case it would be easier to just apply an external static magnetic field? As we will see, this is not the case. Since we take the Floquet photon energy to be considerably larger than the total range of the three energy levels in the static case, the effective Hamiltonian derived from a high-frequency expansion in Sec.~\ref{sec:theory}F will provide a precise answer to this question, among other physical insights, at least for small Floquet amplitudes. 

Interestingly, for general total spin $S$ the non-equilibrium term Eq.~\eqref{ham_neq_hf} will create novel spin couplings that generally do not map onto the standard zero-field and Zeeman terms. For the remainder of the discussion, we will restrict to $S = 1$ and the particular form of the Floquet drive we have applied in this work given by Eq.~\eqref{magnetic_mono_exp}. In that case, the non-equilibrium term reduces to a simpler form (Eq.~\eqref{ham_neq_spin1_hf}), which indicates that the diagonal elements of the zero-field tensor $\tilde{D}$ are renormalized. Furthermore, while we begin with a diagonal $\tilde{D}$ in the static case, the Floquet drive can introduce non-zero off-diagonal terms (Eq.~\eqref{ham_deltazf_hf}). Additionally, even if $\vec{B}_s = 0$ in the static case, an effective Zeeman shift may be generated (Eqs.~\eqref{magnetic_eff_1_hf} and~\eqref{magnetic_eff_22_hf}). 

Given our previous discussion in Sec.~\ref{sec:theory}E, since $\tilde{D}^{\textrm{(eff)}}$ is independent of $\vec{B}_s$, the SMFS and dynamical cancellation problems are equivalent at this level of theory. This means that when $\vec{B}^{\textrm{(eff)}} = 0$ all three renormalized energy levels have SMFS and any pair of distinct levels forms a clock transition. When $\vec{B}^{\textrm{(eff)}} = 0$ the modification of the energy levels by the Floquet drive is completely due to a renormalization of the zero-field tensor. For linear polarization, if $\vec{B}_s = 0$ then we see $\vec{B}^{\textrm{(eff)}} = 0$ because of the cross products in Eqs.~\eqref{magnetic_eff_1_hf}-\eqref{magnetic_eff_22_hf}. For circular polarization, again, due to the cross products in the same equations, we expect $\vec{B}^{\textrm{(eff)}} = 0$ to be solved by some non-zero $\vec{B}_s$. These observations are in general agreement with our numerical simulations, although we emphasize that in reality the solutions to the two problems are slightly different with $\tilde{D}^{\textrm{(eff)}}$ having a corresponding weak dependence on $\vec{B}_s$.

Lastly, independent of $\vec{B}_s$, a drive linearly polarized along one of the principal axes causes the off-diagonal corrections to $\tilde{D}$ to be zero ($C_{1m} = C_{2m} = C_{3m} = 0$), and only one of the coefficients that goes into the diagonal renormalization terms can be non-zero at most (see $C_{4m}$, $C_{5m}$, and $C_{6m}$). If one further restricts the control to the $z$ principal axis with $E = 0$, then all six $C_{\ell m}$ coefficients are zero and the Floquet drive has no effect on $\tilde{D}$. In that case, if $\vec{B}_s = 0$ as well so that $\vec{B}^{\textrm{(eff)}} = 0$, then the effective Hamiltonian is the static Hamiltonian. This means the Floquet drive would have no effect on any of the three energy levels, which is consistent with the top right panel of Fig.~\ref{fig:energy_levels_ED0}. A more simple explanation for this case is that the time-dependent Hamiltonian itself commutes with $s_z$.


\subsection{Realization of the Floquet Control}

The experimental realization of the proposed system will require producing oscillating magnetic fields with amplitudes up to a few hundred mT at photon energies of 20 $\mu$eV, corresponding to oscillation frequencies of around 5 GHz. Perhaps the simplest idea is to pass an oscillating current through a straight wire or coil, although the required current amplitude may be prohibitively large. Alternatively, one could drive a high-Q superconducting resonant cavity with a klystron or similar source. 

A more localized approach relies on generating magnetic moments in a material via optical excitation, with one mechanism being the inverse Faraday effect \cite{van1965optically, pershan1966theoretical, hertel2006theory, kurkin2008transient, de2012coherent, battiato2014quantum, davies2020pathways, zhang2022all, khokhlov2024double, hennecke2024ultrafast}. We take the atomically thin ferromagnetic semiconductor CrI$_3$ as an example \cite{zhang2022all}. A 100 fs optical pulse at 2.18 eV with a fluence of 50 $\mu$J/cm$^2$ enables a complete switching of the magnetization, with the direction determined by the helicity. Applying a train of such pulses with alternating helicities and the correct time delay would then produce an oscillating magnetization at the desired frequency \footnote{If the resulting time-dependent magnetic field at the magnetic center is time-periodic but not monochromatic, our Floquet analysis will still apply as long as the high-order Fourier components are included in Eq.~\eqref{ham_fourier}}. While they switch at 80 MHz in this work, there are no fundamental barriers to increasing this frequency to our targeted value of 5 GHz and far beyond, given that the required pulse lengths are only 100 fs. Another experiment demonstrated, in the absence of switching, that the effective magnetic field experienced by WSe$_2$ sitting directly on the surface of CrI$_3$ can be greater than 10 T \cite{zhong2017van}. This sets an upper limit for the Floquet amplitude that can be generated at the magnetic center of a molecule sitting on the same surface, and is considerably larger than the maximum value considered in our simulations (300 mT). By initially demagnetizing the sample and tuning the fluence below the above stated threshold for complete magnetization switching, the Floquet control can be realized at the desired frequency and amplitude. Alternatively, or perhaps in conjunction, adding buffer layers of insulating and non-magnetic hexagonal boron nitride can reduce the effective magnetic field amplitude.

Moreover, all approaches produce a linearly polarized drive at the magnetic center, which will trivially generate renormalized energy levels with SMFS at $\vec{B}_s = 0$. Practically, one may chemically tune the molecule as close as possible to the desired energy level structure and then use relatively low amplitude Floquet drives for further precision tuning.


\subsection{Operation of Floquet Qubits}

While our focus has been on the renormalization of the quasienergy levels, for quantum information applications a consideration of the corresponding quantum states is essential. In particular, the computational basis for traditional qubits consists of a pair of eigenstates from a time-independent Hamiltonian. In our case, because of the external time-periodic magnetic field, the quantum states are now dynamical, raising questions about the system's potential use as a qubit. Fortunately, previous work has shown that a pair of Floquet eigenstates can in fact be utilized as a computational basis, thus forming a \textit{Floquet qubit} \cite{huang2021engineering, gandon2022engineering}. In particular, it was demonstrated that state initialization, single- and two-qubit gate operations, and state readout can be executed with fidelities exceeding 99.95\%. It is important to note, however, that while these previous studies focused on superconducting qubits, our work centers on spin qubits.


\section{Conclusions and Outlook}
\label{sec:conc}

In this paper we explored an approach for altering the properties of single metal center magnetic molecules that function as spin qubits. Quantum control by Floquet engineering is realized by coupling the total emergent electronic spin $S$ to a time-periodic magnetic field. We demonstrate significant continuous tunability of the low-lying energy levels, which in practice will be ultrafast and reversible via amplitude modulation of the drive on the appropriate timescales. In particular, we focus on an electronic spin Hamiltonian containing the zero-field and Zeeman terms with $S = 1$, where the three energy levels feature SMFS for zero external static magnetic field (see Eq.~\eqref{smfs_condition}). Our essential finding is that for linear polarizations of the dynamical control, all three energy levels retain their SMFS for all Floquet amplitudes under consideration (up to 300 mT). Physical insight into this behavior is enabled by deriving a high-frequency effective Hamiltonian, and also by comparing to the results of what we term the dynamical cancellation problem. If a pair of states functions as a qubit, retaining SMFS will ensure a suppression of decoherence \cite{shiddiq2016enhancing, kundu2023electron, chen2025enhancing}, while their energy separation can be tuned to an experimentally convenient value. 

In the process, several ideas for future research directions have emerged. While we have focused on $S = 1$, it could be interesting to investigate how the Floquet drive modifies the energy levels of larger emergent total spin manifolds \cite{stoll2006easyspin}. Additionally, more physics could be incorporated into the spin Hamiltonian by including the exchange and hyperfine terms, for instance \cite{stoll2006easyspin}. Although originally motivated by magnetic molecules, our theory applies equally well to other systems featuring total spin $S$ manifolds, such as defects or impurities in solids \cite{liu20192d, chatterjee2021semiconductor, wolfowicz2021quantum, burkard2023semiconductor}. Of direct relevance to the current paper is the $S = 1$ electronic spin of the NV center in diamond \cite{naydenov2011dynamical, herbschleb2019ultra}. Given its negligible $E$ zero-field splitting parameter, applying Floquet driving would open a tunable clock transition gap between the upper two energy levels that is not possible in the static case by simply applying an external static magnetic field. 

Furthermore, going beyond a monochromatic Floquet drive is possible with the current formalism by incorporation of the higher-order Fourier components. In this direction, a quantum optimal control approach could be leveraged to tailor the Fourier components to achieve a desired molecular energy level structure \cite{castro2022floquet, castro2023floquet, castro2024qocttools}. For quantum computing applications, it will be essential to determine how the spin coherence times are modified by the Floquet control, which can be modeled via quantum master equations including spin-phonon interactions and coupling to the nuclear spin bath \cite{yu2022ampere, krogmeier2024low}. Lastly, although prior work has explored Floquet qubit operation in superconducting systems \cite{huang2021engineering, gandon2022engineering}, the procedures for quantum state initialization, gate implementation, and readout must be reexamined in the context of Floquet spin qubits.


\section*{Acknowledgments}

It is a pleasure to thank Xiao Chen (Northeastern University), Alberto de la Torre, Mingzhong Wu, Jeffrey Rable, Stephen Hill, Brendan Sheehan, Joshuah Heath, Alexander Balatsky, and Pavel Volkov for stimulating discussions. This work was supported as part of the Center for Molecular Magnetic Quantum Materials, an Energy Frontier Research Center funded by the U.S. Department of Energy, Office of Science, Basic Energy Sciences under Award No. \mbox{DE-SC0019330}. The authors acknowledge UFIT Research Computing for providing computational resources (HiPerGator) and support that have contributed to the research results reported in this publication. This research used resources of the National Energy Research Scientific Computing Center (NERSC), a Department of Energy Office of Science User Facility using NERSC award \mbox{BES-ERCAP0022828}.


\appendix


\section{Expansion Coefficients of the Exact Numerical Effective Hamiltonian}
\label{sec:app-exact}

The first step is to convert the $3 \times 3$ matrices in Eq.~\eqref{ham_eff_expansion} into $9 \times 1$ vectors
\begin{equation}
\vec{h}^{\textrm{(eff)}}
=
\sum_{b=1}^{9}
c_b
\vec{m}_b
.
\label{ham_eff_expansion_vector}
\end{equation}
Defining the following vectors and matrices
\begin{equation}
\vec{h}^{\textrm{(eff)}}
\equiv
\begin{bmatrix}
(\vec{h}^{\textrm{(eff)}})_1 \\
(\vec{h}^{\textrm{(eff)}})_2 \\
\vdots \\
(\vec{h}^{\textrm{(eff)}})_9
\end{bmatrix} 
,
\label{ham_eff_vector}
\end{equation}
\begin{equation}
M
\equiv
\begin{bmatrix}
(\vec{m}_1)_1 & (\vec{m}_2)_1 & \hdots & (\vec{m}_9)_1 \\
(\vec{m}_1)_2 & (\vec{m}_2)_2 & \hdots & (\vec{m}_9)_2 \\
\vdots & \vdots & \ddots & \vdots \\
(\vec{m}_1)_9 & (\vec{m}_2)_9 & \hdots & (\vec{m}_9)_9 \\
\end{bmatrix}
,
\label{m_matrix}
\end{equation}
\begin{equation}
\vec{c}
\equiv
\begin{bmatrix}
c_1 \\
c_2 \\
\vdots \\
c_9
\end{bmatrix} 
,
\label{c_vector}
\end{equation}
Eq.~\eqref{ham_eff_expansion_vector} can be written as
\begin{equation}
\vec{h}^{\textrm{(eff)}} = M \vec{c}
.
\label{ham_eff_matrix_equation}
\end{equation}
We find that
\begin{equation}
\textrm{det}(M) \neq 0
,
\label{det_m_matrix}
\end{equation}
so that there is only one solution to Eq.~\eqref{ham_eff_matrix_equation}, which is given by
\begin{equation}
\vec{c} = M^{-1} \vec{h}^{\textrm{(eff)}}
.
\label{c_vector_solution}
\end{equation}


\section{Effective Hamiltonian from Van Vleck Expansion}
\label{sec:app-vV}

Within Van Vleck degenerate perturbation theory, the first three terms in the series expansion for the effective Hamiltonian in the high-frequency limit are given as \cite{mikami2016brillouin}
\begin{equation}
H^{\textrm{(eff)}}
=
H_0^{\textrm{(eff)}} + H_1^{\textrm{(eff)}} + H_2^{\textrm{(eff)}} + \cdots
,
\label{vv_expansion}
\end{equation}
\begin{equation}
H_0^{\textrm{(eff)}}
=
H^{(0)}
,
\label{vv0}
\end{equation}
\begin{equation}
H_1^{\textrm{(eff)}}
=
\sum_{m \neq 0}^{}
\frac{ [ H^{(-m)} , H^{(m)} ] }{2 m \hbar \Omega}
,
\label{vv1}
\end{equation}
\begin{align}
H_2^{\textrm{(eff)}}
&=
\sum_{m \neq 0}^{}
\frac{ [ [ H^{(-m)} , H^{(0)} ] , H^{(m)} ] }{2 (m \hbar \Omega)^2}
\notag
\\
&+
\sum_{m \neq 0}^{}
\sum_{n \neq 0,m}^{}
\frac{ [ [ H^{(-m)} , H^{(m-n)} ] , H^{(n)} ] }{3 m n (\hbar \Omega)^2}
,
\label{vv2}
\end{align}
in terms of the Fourier components of the time-dependent Hamiltonian defined in Eq.~\eqref{ham_fourier}.


\section{State Labeling of the Energy Curves: Wave Function Tracking}
\label{sec:app-state}

In the static case ($B_F = 0$), the character of the three energy levels is determined by the values of the zero-field splitting parameters $D$ and $E$. For a given control polarization, each time the Floquet amplitude is incremented, the overlap between the current three and each of the previous three states is computed. Whichever pairs have the largest overlap receive the same label. Mathematically, the overlap is based on the one-cycle average overlap between the appropriate Floquet eigenstates (see Eqs.~\eqref{floquet_ansatz},~\eqref{floquet_eigen_time}, and~\eqref{wavefunction_fourier}) 
\begin{align}
\mathcal{O}_{nn'}(B_F + \delta B_F , B_F)
&\equiv
\notag
\\
\Bigg|
\frac{1}{T}
\int_{0}^{T}
dt
&\Phi_{n , B_F + \delta B_F}^\dagger(t)
\Phi_{n' , B_F}(t)
\Bigg|
,
\label{overlap_floquet_states} 
\end{align}
where $n, n' = 1, 2, 3$. We emphasize that the wave function is not constant over the entire energy curve, but rather indicates a connection to the static state character under adiabatic variation of the Floquet amplitude.


\section{Numerical Search for Energy Level SMFS}
\label{sec:app-clock}

We define an objective function
\begin{equation}
\Theta(\vec{B}_s)
\equiv
\sum_{n=1}^{3}
\left|
\frac{\partial \epsilon_n}{\partial \vec{B}_s}
\right|
,
\label{objective_function}
\end{equation}
which is computed using Eqs.~\eqref{hf_theorem} and~\eqref{hf_matrix_blocks}. The objective function should be minimized to zero if all three energy levels have SMFS at a particular $\vec{B}_s$. In practice, for circular polarization, the objective function will converge to a non-zero value since only one of the three levels will have exactly SMFS for the optimized $\vec{B}_s$ (see Fig.~\ref{fig:energy_sweep_floquet} for a visual example). 

Numerically, we implement an adaptive grid search method. The objective function is computed at a starting vector $\vec{B}_s$ as well as the 26 neighboring points using an initial uniform grid spacing in each of the Cartesian directions of 0.1 mT. Whichever of the 26 grid points produces the lowest value of the objective function becomes the new central reference point for $\vec{B}_s$ in the next iteration. If no shifts lower the objective function, the grid spacing is decreased by a factor of $\sqrt{3}$. 

For a given control polarization, the optimization starts at $B_F = 0$. Each time the Floquet amplitude is incremented, the initial guess for $\vec{B}_s$ is taken to be the vector that was converged for the previous $B_F$. In this way, we find the solutions that are adiabatically connected to the static ($B_F = 0$) solution of $\vec{B}_s = 0$. We do not exclude the possibility that other $\vec{B}_s$ can produce Floquet renormalized energy levels with SMFS. 


\section{Iterative Self-Consistent Procedure for Dynamical Cancellation Problem}
\label{sec:app-cancel}

Comparing Eq.~\eqref{ham_zeeman} with Eq.~\eqref{ham_eff_spin1}, we have the ansatz
\begin{equation}
\vec{\mathcal{B}}^{\textrm{(eff)}}
=
\tilde{g}^T 
\vec{B}^{\textrm{(eff)}}
,
\label{itersc_slashfield}
\end{equation}
or equivalently
\begin{equation}
\vec{B}^{\textrm{(eff)}}
=
(\tilde{g}^T)^{-1} 
\vec{\mathcal{B}}^{\textrm{(eff)}}
,
\label{itersc_field}
\end{equation}
so that in Eq.~\eqref{ham_eff_spin1}
\begin{equation}
\vec{\mathcal{B}}^{\textrm{(eff)}} \cdot \vec{s}
=
(\vec{B}^{\textrm{(eff)}})^T \tilde{g} \vec{s}
.
\label{term_matching_zeeman}
\end{equation}
The iterative self-consistent procedure for solving $\vec{\mathcal{B}}^{\textrm{(eff)}} = 0$ is as follows. For $\tilde{n} = 1, 2, 3, ...$ perform the update
\begin{equation}
\vec{B}_{s,[\tilde{n}]}
=
\vec{B}_{s,[\tilde{n}-1]}
-
\vec{B}_{[\tilde{n}-1]}^{\textrm{(eff)}}
.
\label{itersc_update1}
\end{equation}
Then use the approach in Sec.~\ref{sec:theory}E to compute $\vec{\mathcal{B}}_{[\tilde{n}]}^{\textrm{(eff)}}$ numerically. If $|\vec{\mathcal{B}}_{[\tilde{n}]}^{\textrm{(eff)}}| < 10^{-4}$ mT, then we consider the algorithm to have converged. Otherwise, we continue with
\begin{equation}
\vec{B}_{[\tilde{n}]}^{\textrm{(eff)}}
=
(\tilde{g}^T)^{-1} 
\vec{\mathcal{B}}_{[\tilde{n}]}^{\textrm{(eff)}}
,
\label{itersc_update2}
\end{equation}
and go back to Eq.~\eqref{itersc_update1}. The only exception to the above general recipe is that we should initialize $\vec{B}_{s,[1]}$ in the same way that was described for the SMFS problem in the previous Appendix. Note that in this Appendix we use the notation $[\cdots]$ to indicate that the subscript index is counting the iteration number in the numerical algorithm, whereas in Eqs.~\eqref{magnetic_eff_1_hf}-\eqref{magnetic_eff_22_hf} the subscript indices are labeling the effective magnetic fields originating from the different orders in the Van Vleck expansion.

Concretely, for a given control polarization, the optimization starts at $B_F = 0$. Each time the Floquet amplitude is incremented, the initial guess for $\vec{B}_s$ is taken to be the vector that was converged for the previous $B_F$. In this way, we find the solutions that are adiabatically connected to the static ($B_F = 0$) solution of $\vec{B}_s = 0$. We do not exclude the possibility that other $\vec{B}_s$ can solve the equation $\vec{\mathcal{B}}^{\textrm{(eff)}} = 0$.

This procedure is self-consistent in the sense that upon convergence, repeated iterations will continue to produce the same static magnetic field solution to within some tolerance.


\FloatBarrier

\begin{table*}[!htb]
\centering
\caption{\label{tab:parameters}Spin and Floquet control parameters utilized in this work.}
\begin{tabular}{cc}
\toprule
Parameter & \hspace{4em}Values \\
\midrule
Total electronic spin $S$ & \hspace{4em}1 \\
Zero-field splitting $D$ & \hspace{4em}5 $\mu$eV  \\
Zero-field splittings $E/D$ & \hspace{4em}0, 0.1, 1/3 \\
Isotropic $g$-tensor value & \hspace{4em}2 \\
Floquet linear polarizations & \hspace{4em}$x$, $y$, $z$ \\
Floquet tilted linear polarizations & \hspace{4em}$+x \pm y$, $+x \pm z$, $+y \pm z$ \\
Floquet circular polarizations & \hspace{4em}$(xy)\pm$, $(xz)\pm$, $(yz)\pm$ \\
Floquet photon energy & \hspace{4em}20 $\mu$eV \\
Floquet amplitudes & \hspace{4em}0-300 mT \\
\bottomrule
\end{tabular}
\end{table*}

\FloatBarrier


\newpage

\bibliography{references}


\end{document}